\documentclass[twocolumn,showpacs,aps,superscriptaddress]{revtex4-1}
\usepackage{color}
\usepackage{soul}
\definecolor{darkgreen}{rgb}{0,0.6,0}
\definecolor{orange}{rgb}{0.99,0.257,0}
\usepackage{hyperref}

\usepackage{graphicx}
\usepackage{amsmath}
\usepackage{amssymb}

\newcommand{\be}{\begin{equation}}
\newcommand{\ee}{\end{equation}}
\newcommand{\ba}{\begin{eqnarray}}
\newcommand{\ea}{\end{eqnarray}}
\newcommand{\bi}[1]{Fig.~\ref{fig:#1}}
\newcommand{\e}[1]{eq.~(\ref{eq:#1})}

\usepackage{lipsum}
\usepackage{layouts}

\begin{document}

\title{An analytical approach to the Mean--Return-Time~Phase of isotropic stochastic oscillators}

\author{Konstantin~Holzhausen}
\affiliation{Bernstein Center for Computational Neuroscience Berlin, Philippstr.~13, Haus 2, 10115 Berlin, Germany}
\affiliation{Physics Department of Humboldt University Berlin, Newtonstr.~15, 12489 Berlin, Germany}

\author{Peter~J.~Thomas}
\affiliation{Department of Mathematics, Applied Mathematics and Statistics, 212~Yost~Hall, Case Western Reserve University, 10900~Euclid Avenue, Cleveland, Ohio, USA}

\author{Benjamin~Lindner}
\affiliation{Bernstein Center for Computational Neuroscience Berlin, Philippstr.~13, Haus 2, 10115 Berlin, Germany}
\affiliation{Physics Department of Humboldt University Berlin, Newtonstr.~15, 12489 Berlin, Germany}
\date{\today}

\begin{abstract}
One notion of phase for stochastic oscillators is based on the mean return-time (MRT): a set of points represents a certain phase if the mean time to return from any point in this set to this set after one rotation is equal to the mean rotation period of the oscillator (irrespective of the starting point). For this so far only algorithmically defined phase, we derive here analytical expressions for the important class of isotropic stochastic oscillators. This allows us to evaluate cases from the literature explicitly and to study the behavior of the MRT phase in the limits of strong noise. We also use the same formalism to show that lines of constant return time variance (instead of constant mean return time) can be defined, and that they in general differ from the MRT-isochrons.
\end{abstract}	
 
\maketitle

Oscillatory processes encountered in physics and biology often come along with a substantial component of randomness and are then termed \emph{stochastic oscillations}. Intracellular calcium waves \cite{SkuKet08}, neural action potentials \cite{WalAlv06}, rhythmic fluctuations  in lasing intensity \cite{McNWie88}, and population oscillations of prey-predator systems \cite{McKNew05} all show pronounced fluctuations in their amplitude as well as in the timing of the cycles.

In order to find a reduced description of such random oscillations, in recent years efforts have been made to generalize the phase concept for deterministic oscillators \cite{Win74,Guc75} (which has also been applied to systems with weak noise, see e.g.~\cite{FreNei00,FreSch03,ZhoBur13,MaKli14}) to the stochastic case \cite{SchPik13} (see also \cite{ThoLin14,GiaPoqSha2018,AmiHolSri2019,EngKue2021}).  
The first proposal, the mean--return-time (MRT) phase by Schwabedal and Pikovsky \cite{SchPik13}, is an intuitive generalization of the stroboscopic return-time-phase definition for deterministic oscillators: in the two-dimensional case, with any starting point on the  line of equal phase (isochron), the mean return-time to the very same line after one rotation is equal to the mean rotation period of the oscillator. Schwabedal and Pikovsky suggested an algorithmic procedure to deform a line until (experimental or simulated) data satisfy this criterion. Cao et al.~\cite{CaoLin20} showed more recently, that the MRT phase  for planar white-Gaussian noise driven oscillators obeys a partial differential equation with an unusual jump condition. 

Based on \cite{CaoLin20} we  present an analytical solution for the MRT phase  for the important class of isotropic stochastic oscillators. 
The solution is given in terms of quadratures and is exploited to explore the limit of strong noise and the role of boundaries for so-far purely numerically treated examples from the literature. 
Finally, our approach allows us to go beyond the familiar notion of isochrons (lines of constant mean rotation time) to introduce a new construct, the \emph{iso-variance} lines (lines of constant variance of the rotation time). We  
demonstrate that the iso-variance lines strongly differ from the isochrons.  

\textit{Derivation~of~the~isochron~expression.--} We consider a class of rotationally invariant planar stochastic oscillators (also called isotropic stochastic oscillators) that obey
\begin{equation}
	\begin{aligned}
		\dot{\rho} &= g(\rho) + q_{\rho}(\rho) \, \xi_{\rho}(t) \\
		\dot{\phi} &= f(\rho) + q_{\phi}(\rho) \, \xi_{\phi}(t)
	\end{aligned}
	\label{eq:GeneralModel}
\end{equation}
in polar coordinates, where $\xi_{\rho, \phi}$ denote white Gaussian noise sources with $\left\langle \xi_{i}(t)\, \xi_{j}(t^{\prime}) \right\rangle = \delta_{i, j}\, \delta(t - t^{\prime}), \; i, j \in \{\rho,\, \phi\}$. 
If the noise is multiplicative ($q_{\rho}(\rho)\neq\text{const}$) we interpret it in the sense of Ito \cite{Gar85}. We impose reflecting boundaries at an inner circle with radius $\rho_-$ and an outer circle $\rho_+$ such that the dynamics is restricted to $\rho_-<\rho(t)<\rho_+$. We assume that there is a mean rotation around the origin, the rotational sense  of which may change depending on the parameters. 

Because none of the functions on the right-hand side depend explicitely on the phase $\phi$, the stochastic dynamics is rotationally invariant, which is why we refer to  \e{GeneralModel}   as an isotropic oscillator. 
In particular, we expect that the isochrons for different phases $\psi\in[0,2\pi]$  will be rotationally invariant as well, and can be expressed via $\varphi(\rho)=\phi_\text{I}(\rho)+\psi$. In the following we  calculate $\phi_\text{I}(\rho)$.

According to \cite{CaoLin20}, the isochrons are the level sets of the mean-first-passage-time function $T(\rho,\, \phi)$ that is the solution to the partial differential equation
\begin{equation}
	\mathcal{L}^{\dagger}\, T(\rho,\, \phi) = \left[ g \partial_\rho+  f\partial_\phi+\frac{q_{\rho}^2}{2}\partial_\rho^2+\frac{q_{\phi}^2}{2}\partial_\phi^2\right] T(\rho,\, \phi)= -1 \, 
	%\quad \text{on} \quad \left[ \rho_{-}, \, \rho_{+} \right] \times \mathbb{R}
	\label{eq:pde}
\end{equation}
satisfying the periodic-plus-jump condition at an arbitrary connection between the inner and outer boundaries (here we choose for simplicity the line $\phi=0$ and consider a counterclockwise rotation):
\begin{equation}
	T(\rho,\, 2 \pi ) + \overline{T} = T(\rho, 0) 
	\label{eq:PeriodicPlusJumpCond}
\end{equation}
and adjoint reflecting boundary conditions at $\rho_\pm$.
%\begin{equation}
%	\partial_{\rho}\,\left. T(\rho,\, \phi) \right|_{\rho_{-},\, \rho_{+}} = 0 \text{.}
%	\label{eq:AdjReflBC}
%\end{equation}
$\overline{T}$ denotes the mean period, the time it takes one realization of the process on average to perform one revolution around the origin. 
%For oscillators, this usually means one revolution in a neighbourhood of a stable limit cycle. 
Following \cite{CaoLin20}, we assume $f, g, q_{\rho}, q_{\phi} \in \mathcal{C}^{2}$ as well as $q_{\rho}^{2}(\rho) > 0$ and $q_{\phi}^{2}(\rho) > 0$ for all $\rho \in [\rho_{-},\, \rho_{+}]$.  The choice of polar coordinates translates the rotational invariance of \e{GeneralModel} into a translational invariance in the angular coordinate $\phi$. Correspondingly, a symmetry-adapted solution is of the general form
\begin{equation}
	T(\rho,\, \phi) = -\frac{\overline{T}}{2\, \pi} \, \phi + T_{\rho}(\rho)
	\label{eq:SymmAdapAnsatz}
\end{equation}
that also fulfills the periodic-plus-jump condition. This symmetry-adapted ansatz reduces the number of dimensions in \e{pde}, giving  an ordinary differential equation for $T_{\rho}$ in $\rho$, which can be solved explicitly in terms of quadratures. The isochron $\phi_{\textrm{I}}$ as a special level set of  $T(\rho, \phi)$ can then be parametrized according to $\phi_{\textrm{I}}(\rho) = 2 \pi T_{\rho}(\rho)/\overline{T}$, leading to our main result: 
\begin{equation}
\phi_{\textrm{I}}(\rho) \!=\! 2\!\int\limits_{\rho_{-}}^{\rho}\! dq\! \int\limits_{\rho_{-}}^{q}\! du \frac{f(u) - \overline{\omega}}{q_{\rho}^{2}(u)} 
\exp\left[{-2 \int_{u}^{q} dv \frac{g(v)}{q_{\rho}^{2}(v)}}\right].
	\label{eq:IsoAngParam}
\end{equation}
The mean rotation frequency $\overline{\omega}$ (essentially, the inverse mean rotation period $2\pi/\bar{T}$) is given by
\begin{equation}
%	\begin{aligned}
		\overline{\omega} = \frac{2\pi}{\overline{T}}= \frac{\int_{\rho_{-}}^{\rho_{+}} d\rho \, f(\rho) / q_{\rho}^{2}(\rho)\,
			e^{-2 \int_{\rho}^{\rho_{+}} d\rho^{\prime} q_{\rho}^{-2}(\rho^{\prime})\, g(\rho^{\prime})}}
		   {\int_{\rho_{-}}^{\rho_{+}} d\rho\, q_{\rho}^{-2}(\rho)\,
			e^{-2 \int_{\rho}^{\rho_{+}} d\rho^{\prime} q_{\rho}^{-2}(\rho^{\prime})\, g(\rho^{\prime})}}.
%	\end{aligned}
%	\quad \text{.}
	\label{eq:MeanAngFreq}
\end{equation}
We can make a few conclusions from the analytical result \e{IsoAngParam} without any numerical evaluation: 
i) $q_{\phi}$, i.e.~the strength of phase diffusion, does not  enter the isochron parametrization at all (all oscillators that only differ with respect to $q_{\phi}$ have the same isochrons);
ii) the slope of the phase $d\phi/d\rho$ is given by the  difference between the local phase progression speed $f(\rho)$ and the average speed $\overline{\omega}$ weighted with the (normalized) steady state density of the radius and averaged over the interval $(\rho_-,\rho)$; 
iii) if the overall strength of the radial noise increases without bounds and we keep the two boundary values $\rho_\pm$ fixed at non-vanishing finite values, the integrand in \e{IsoAngParam} approaches zero, i.e.~the isochron $\phi_\textrm{I}$ does not depend on the radius anymore and isochrons %attend 
approach the shape of spokes of a wheel. 

\emph{Newby-Schwemmer oscillator.--} This is given by \cite{NewSch14}
\begin{equation}
    \begin{aligned}
        \dot{\rho} &= -\gamma\, \rho\, (\rho^{2} - 1) + \frac{D}{\rho} + \sqrt{2\, D} \, \xi_{\rho}(t),\\
        \dot{\phi} &= \omega + \omega\, \gamma\, c\, (1 - \rho)^{2} + \frac{\sqrt{2\, D}}{\rho} \, \xi_{\phi}(t).
    \end{aligned}
    \label{eq:NewbySchwemmer}
\end{equation}
The sense of rotation is counterclockwise on the limit cycle at $\rho=1$ but, for $c<0$, turns to clockwise %counterclockwise 
when the trajectory deviates sufficiently both towards the inside and the outside. 
With noise, the occupation probability w.r.t.~$\rho$ determines whether the net rotation goes clockwise or counterclockwise \cite{NewSch14}, and at the transition point $D=D_t$ between both regimes, the mean rotation time diverges (see \bi{NewbySchwemmer}c)  

\begin{figure}[h]
    \centering
    \includegraphics[width=2.7in]{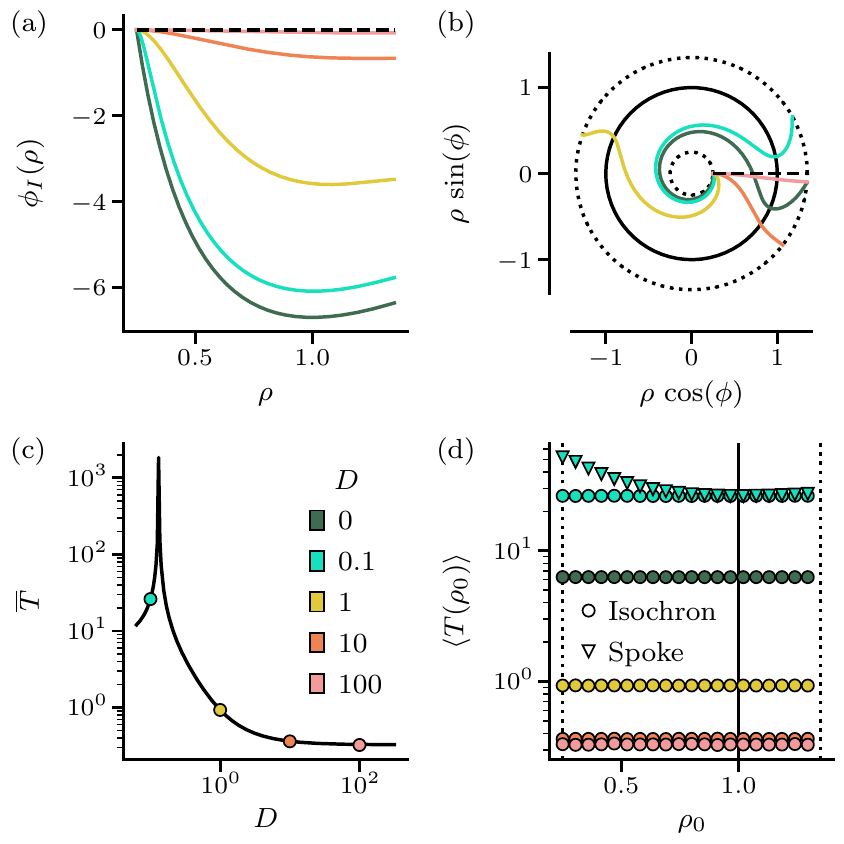}
    \caption{{\bf MRT-Isochrons of the Newby-Schwemmer model.} (a): Isochron parametrization $\phi_\textrm{I}(\rho)$, \e{IsoAngParam}. (b): Isochrons (colored), reflecting boundaries (dotted) and limit cycle (solid black). (c): Mean period $\overline{T}$ vs $D$ according to theory (lines) and simulations (circles).
     (d): Mean return-time  $\left\langle T(\rho_0) \right\rangle$ to a curve vs initial position $\rho_0$ on this curve, for the curve being an isochron (circles, different colors represent different $D$) or a  spoke (triangles, here $D=0.1$).}\label{fig:NewbySchwemmer}
\end{figure}

The deterministic isochron (dark green line in \bi{NewbySchwemmer}a and b) has a hook-like shape, because points away from the limit cycle need some head start (net rotation is counterclockwise). 
For non-vanishing noise this principal shape is maintained: for $D<D_t$ the net rotation is still counterclockwise, and points off the limit cycle need a headstart; for $D>D_t$  these points need a ``later" start (net rotation is clockwise), thus the isochron has the same curvature. 
In addition, with increasing noise intensity $D$ the isochrons become flatter, approaching the spoke of a wheel (dashed black line) as predicted by our general conclusion (iii) above.

We checked the constant-MRT property by performing extensive simulations with ensembles of stochastic trajectories for a certain noise intensity $D$  starting on the corresponding isochron. Indeed, the MRT for a given noise intensity from and to the isochron do not depend on the specific starting point along the isochron (circles in \bi{NewbySchwemmer}d). In marked contrast, choosing (for a comparatively small noise intensity) as initial set and target line the spoke of a wheel, the mean return times \emph{do} depend on the initial position (triangles in \bi{NewbySchwemmer}d). 

\textit{Guckenheimer-Schwabedal-Pikovsky oscillator.--} A classic example of a deterministic system, in which the definition of phase is problematic,  is one with two limit cycles with distinct rotation frequencies (this example goes back to Guckenheimer \cite{Guc75}). As shown numerically by Schwabedal and Pikovsky \cite{SchPik13}, with noise, transitions occur between the attraction domains of the two limit cycles and thus the MRT phase uniquely defines one phase for the entire system. This system is given by 
\begin{equation}
    \begin{aligned}
        \dot{\rho} &= \rho\, (1-\rho)(3-\rho)(c-\rho) + \frac{\sigma^{2}\, \rho}{2} + \sigma\, \rho \xi(t) \\
        \dot{\phi} &= \omega + \delta\, (\rho - 2) - (1 - \rho)\, (3 - \rho)
    \end{aligned}
    \label{eq:SchwabedalPikovsky}
\end{equation}
\begin{figure}[h]
    \centering
    \includegraphics[width=2.7in]{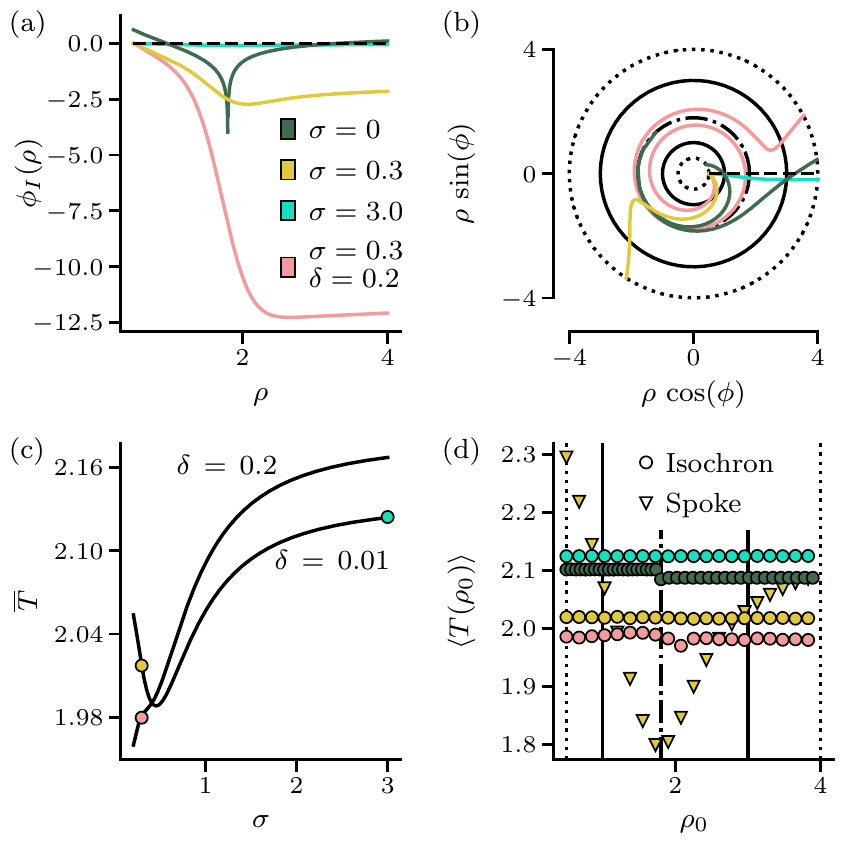}
    \caption{{\bf MRT-Isochrons of the Guckenheimer-Schwabedal-Pikovsky oscillator.} Panels as in \bi{NewbySchwemmer}  but for \e{SchwabedalPikovsky}. Parameters: $\omega=3, c=1.8$, $\rho_-=0.5, \rho_+=4.0$.
 }\label{fig:SchwabPikov}
\end{figure}

 The unstable limit cycle  at $\rho=c$ separates the basins of attraction at $\rho=1$ and $\rho=3$.  With noise, the unstable limit cycle 
 can be surmounted,  smoothing the discontinuity, and thus resulting in one stochastic isochron that is continuous over the whole domain connecting both basins of attraction (see \bi{SchwabPikov}, a,b). 
 
While MRTs for a given noise intensity remain constant for initial positions along the corresponding isochron, they vary significantly if the initial set and target line is a ray emanating from the origin (cf. \bi{SchwabPikov}d). 
For $D=0$, the return time suffers a jump by $2\, \delta$ between the basins of attraction  (i.e.~for an initial point $\rho=c$), reflecting the above mentioned fact that there is no unique phase for the deterministic system. The stronger this jump is, the more twisted the stochastic MRT isochron becomes (cf.~lines for $\delta=0.01$ and $\delta=0.2$ in \bi{SchwabPikov}a,b); $\delta$  may also change how the mean rotation period depends on the noise level (cf.~\bi{SchwabPikov}c) We note that for strongly increasing noise, the shape of the isochron flattens  as again expected from our general consideration.
 
\begin{figure}[h]
    \centering
    \includegraphics[width=2.7in]{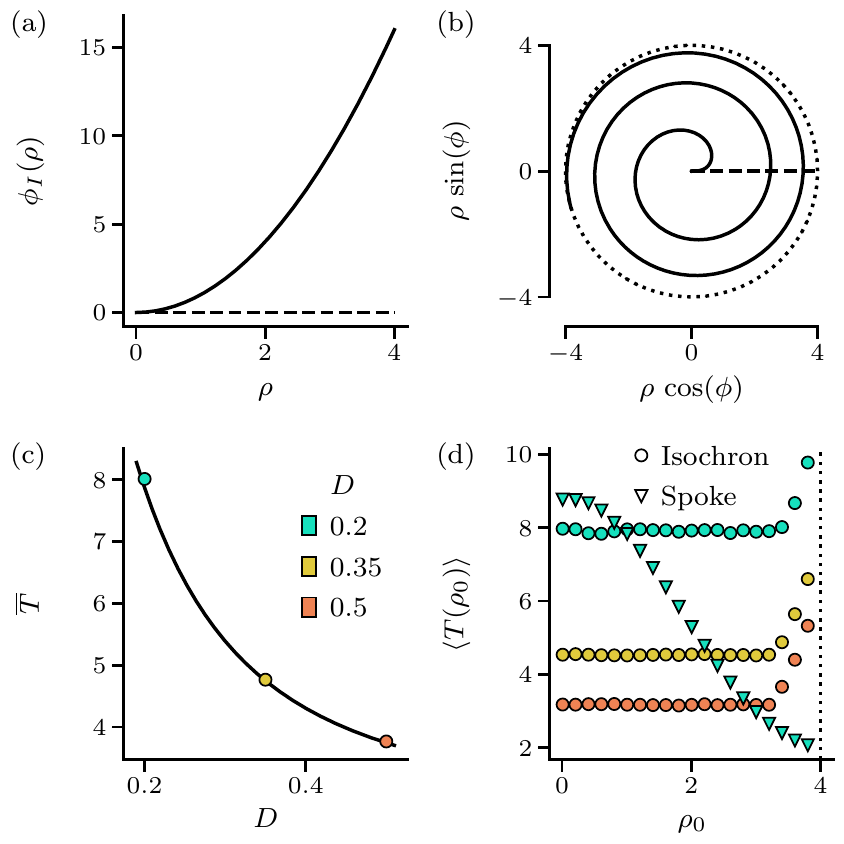}
    \caption{{\bf MRT-isochrons of the monomial model.} 
  Presentation in the panels is as in \bi{NewbySchwemmer} but for \e{SimpleModel} with $n=1, \rho_-=0, \rho_+=\infty$. Simulations in (d) are, however, performed with $\rho_-=10^{-4}$ $\rho_+=4.0$ consequently, the MRT-property is only obeyed for initial points far from the outer boundary. 
  } \label{fig:SimpleModel}
\end{figure}

\textit{A~monomial~model.--} According to what we have found so far, we may expect, that in the strong-noise limit (until now, the less explored regime of stochastic oscillators), the isochrons always approach a radial spoke  
(conclusion (iii) above). This, however, is only true for systems with finite boundaries. 
To show this, we consider   a particularly simple 
model system with monomial frequency dependence, for which the integrals in \e{IsoAngParam} can be expressed by elementary functions. The model obeys
\begin{equation}
    \begin{aligned}
        \dot{\rho} &= -\alpha + \frac{D}{\rho} + \sqrt{2\, D}\, \xi_{\rho}(t) \\
        \dot{\phi} &= -\beta\, \rho^{n} + \frac{\sqrt{2\, D}}{\rho} \, \xi_{\phi}(t)
    \end{aligned}
    \label{eq:SimpleModel}
\end{equation}
with positive parameters $\alpha, \beta$ and integer $n$, resulting in clockwise rotations with a speed that increases as a power law with increasing radius.  We are specifically interested in the limit in which the inner boundary approaches the origin and the outer boundary goes to infinity. In this limit, the MRT and the  isochrons can be expressed by elementary functions \cite{Hol21}; particularly simple and striking is the expression for $n=1$:
\begin{equation}
    \lim_{\substack{\rho_{-} \rightarrow 0 \\ \rho_{+} \rightarrow \infty \\ \alpha > 0}} \phi(\rho) = \frac{\beta}{\alpha} \,\frac{\rho^{2}}{2},\quad \lim_{\substack{\rho_{-} \rightarrow 0 \\ \rho_{+} \rightarrow \infty \\ \alpha > 0}} \overline{T}=\frac{\alpha}{2\beta D}
    \label{eq:isochron_simple}
\end{equation}
Hence, quite surprisingly, the MRT phase is completely independent of the noise level but clearly different from a simple ray
(see \bi{SimpleModel}a,b). Specifically in the limit $D\to \infty$ 
the isochrons do not converge to the spokes of wheel, which is due to the absence of finite boundaries: for stronger and stronger noise, the main share of probability moves to larger and larger radii and the mean rotation time drops strongly (see \bi{SimpleModel}c).

We can test how important the infinite boundary condition is by running stochastic simulations with initial set and target line with the isochron \e{isochron_simple} but imposing a reflecting boundary at a finite value $\rho_+<\infty$ (for this setting, \e{isochron_simple} is \emph{not} the exact isochron). For all noise intensities, the MRT is flat as a function of the initial radius except for a finite region close to the outer boundary (see \bi{SimpleModel}d). 
\begin{figure}[h]
    \centering
    \includegraphics[width=2.7in]{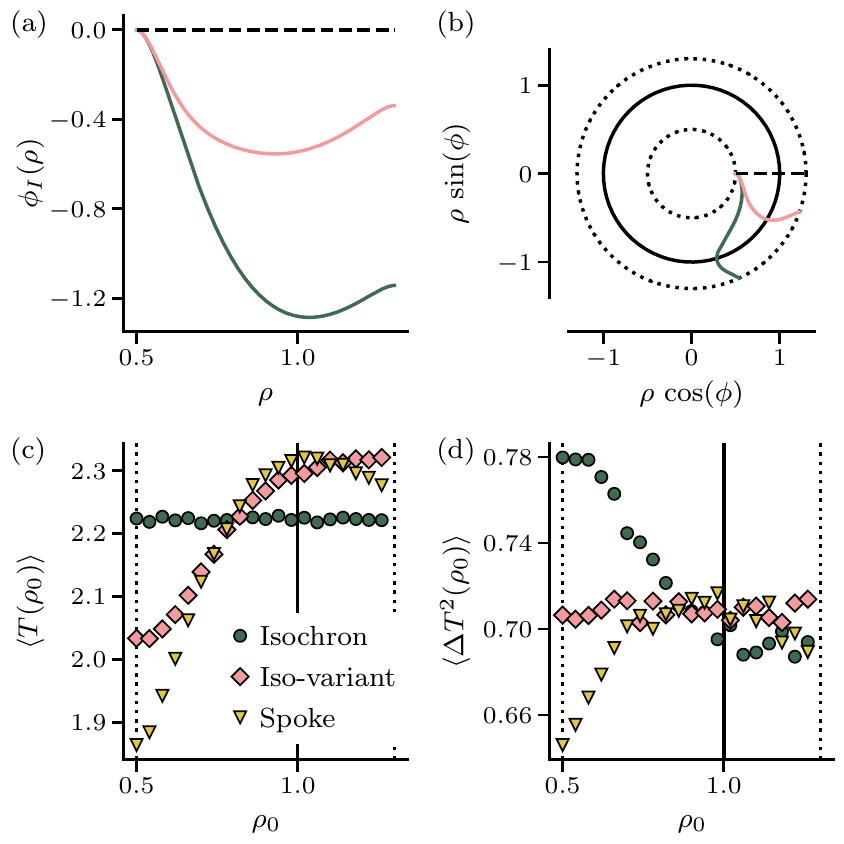}
    \caption{{\bf Iso-variance curves for the Newby-Schwemmer oscillator.}(a): Isochronal curve  and curve of constant return-time variance of the Newby-Schwemmer model in terms of  angle variable vs radius. (b) Same curves as in (a),  in cartesian coordinates.  (c): Mean return time to isochrons, iso-variance line, or spoke as function of the initial point  on the respective line. (d) Variances of the Return times $\left\langle \Delta^{2}\right\rangle>$ as function of initial position on the respective lines.}\label{fig:MeanVarianceComp}
\end{figure}

\textit{Iso-variance curves.--} The same approach used for the isochrons permits to calculate the lines of constant variance, that is, the lines for which the return time to the line after one rotation possesses the same variance irrespective of the starting point. It turns out that these lines are the contour lines of the variance function, obeying
\be
L^\dagger \Delta = q_{\rho}^{2}\, \left( \partial_{\rho}\, T(\rho,\, \phi) \right)^{2} + q_{\phi}^{2}\, \left( \partial_{\phi}\, T(\rho,\, \phi) \right)^{2}
\ee  
with a jump condition $\Delta(\rho,\, \phi) = \Delta(\rho,\, \phi + 2\, \pi) + \overline{\Delta}$ which can be solved for the isotropic oscillator by a similar ansatz $\Delta(\rho)=\Delta_\rho(\rho)-\phi \bar{\Delta}/(2\pi)$ resulting in the parametrization of the iso-variance curve
\be
\begin{aligned}
\phi_\Delta(\rho) &= 2 \int\limits_{\rho_{-}}^{\rho} \!\!dq \int\limits_{\rho_{-}}^{q}\! du  \left[ \frac{2 \pi}{\overline{\Delta}} \left( q_{\rho}^{2}(u) \left( \partial_{u} T_{\rho}(u) \right)^{2} + \frac{q_{\phi}^{2}(u)\, \overline{T}^{2}}{4 \pi^{2}} \right) \right. \\ 
& \left. \frac{}{} - f(u) \right]\, \exp{\left( -2 \int\limits_{u}^{q} dv\, \frac{g(v)}{q_{\rho}^{2}(v)} \right)}\Big/q_{\rho}^2(u)
\end{aligned}
\ee
Here the variance of the rotation time $\bar{\Delta}$ is given by 
\be
\bar{\Delta} = - 2 \pi \frac{\int\limits_{\rho_{-}}^{\rho_{+}} d\rho\, \left[ \left( \partial_{\rho} T_{\rho}(\rho) \right)^{2} + \frac{q_{\rho}^{2}}{q_{\phi}^{2}} \frac{\overline{T}^{2}}{4 \pi^{2}} \right]\, \exp\left(-2 \int\limits_{\rho}^{\rho_{+}} du\,  \frac{g(u)}{q_{\rho}^{2}(u)}\right)}{\int\limits_{\rho_{-}}^{\rho_{+}} d\rho\, f(\rho) \exp\left(-2 \int\limits_{\rho}^{\rho_{+}} du\,  \frac{g(u)}{q_{\rho}^{2}(u)}\right)/q^{2}_{\rho}(\rho)}
\ee
We have evaluated and tested these formulas for the Newby-Schwemmer oscillator \e{NewbySchwemmer} and also compare the resulting iso-variance lines to the isochrons and the spoke of wheel in \bi{MeanVarianceComp}. We observe that isochron and iso-variance curve do not agree in this case: the iso-variance curve (pink line in \bi{MeanVarianceComp}a,b) is significantly less twisted than the isochron (green line). In panels c,d we test the defining properties of the two lines: the MRT is approximately independent of the starting point only for the isochron but neither for the iso-variance curve nor the spoke (c); the variance of the return time is approximately flat as a function of the starting point for the iso-variance line but neither for the isochron nor the spoke. 

\emph{Conclusions.--} We have found the analytical mapping for Schwabedal and Pikovsky's mean return-time (MRT) phase  for the important class of planar isotropic stochastic oscillators and have tested it for three examples. We have seen that for systems with finite boundaries, the phase description in the strong noise limit always yields the geometric phase (spokes of a wheel); for small to moderate noise, the curvature of the MRT phase isochrons reflect the stochastic interplay between radial and angular dynamics.
We also demonstrated that constancy of the first cumulant (the MRT) over the isochron does not imply constancy of higher cumulants, for instance, the variance of the return time for different initial positions on the isochron. We could find an equation and solve it for the iso-variance curve: in our example the resulting line was quite different from the MRT isochron. 

We hope that our results can be used to address important open issues in the analysis of stochastic oscillators: i) the comparison to the asymptotic phase \cite{ThoLin14} in some analytically tractable cases (see e.g. \cite{ThoLin19}); ii) the oscillator response to external perturbations via a generalized phase-response curve \cite{GalErm05,ErmGal07};
iii the analysis of variability in neuronal firing patterns \cite{PuTho2021}; iv) the treatment of networks of strongly stochastic oscillators \cite{ErSau06}.  

%\bibliography{ALL_19_03_11}
%\bibliography{ALL_20_02_26}

%\bibliographystyle{unsrt}
%\bibliography{ALL_21_08_05}
\end{document}